\documentclass[10pt,twocolumn,preprint]{sigplanconf}

\usepackage{amsmath}
\usepackage{epsfig}

\begin{document}

\conferenceinfo{WXYZ '05}{date, City.} 
\copyrightyear{2005} 
\copyrightdata{[to be supplied]} 


\title{On the Practicality of `Practical' Byzantine Fault Tolerance}

\authorinfo{Nikos Chondros}
           {University of Athens}
           {n.chondros@di.uoa.gr}
\authorinfo{Konstantinos Kokordelis}
           {University of Athens}
           {kokordelis.konstantinos@gmail.com}
\authorinfo{Mema Roussopoulos}
           {University of Athens}
           {mema@di.uoa.gr}

\maketitle

\begin{abstract}
Byzantine Fault Tolerant (BFT) systems are considered by the systems 
research community to be state of the art with regards to providing 
reliability in distributed systems. BFT systems provide safety and liveness 
guarantees with reasonable assumptions, amongst a set of nodes where at 
most f nodes display arbitrarily incorrect behaviors, known as Byzantine 
faults. Despite this, BFT systems are still rarely used in practice. In 
this paper we describe our experience, from an application developer's 
perspective, trying to leverage the publicly available and highly-tuned 
``PBFT'' middleware (by Castro and Liskov), to provide provable reliability 
guarantees for an electronic voting application with high security and 
robustness needs. The PBFT middleware has been the focus of most BFT 
research efforts over the past twelve years; all direct descendent systems 
depend on its initial code base.

We describe several obstacles we encountered and drawbacks we identified 
in the PBFT approach. These include some that we tackled, such as lack 
of support for dynamic client management and leaving state management 
completely up to the application. Others still remaining include the 
lack of robust handling of non-determinism, lack of support for web-based 
applications, lack of support for stronger cryptographic primitives, and 
others. We find that, while many of the obstacles could be overcome with 
a revised BFT middleware implementation that is tuned specifically for 
the needs of the particular application, they require significant engineering 
effort and time and their performance implications for the end-application 
are unclear. An application developer is thus unlikely to be willing to 
invest the time and effort to do so to leverage the BFT approach. We conclude 
that the research community needs to focus on the usability of BFT algorithms 
for real world applications, from the end-developer perspective, in addition 
to continuing to improve the BFT middleware performance, robustness and 
deployment layouts.

\end{abstract}




\section{Introduction}
\label{sect:intro}

Byzantine Fault Tolerant (BFT) systems are considered by the systems
research community to be state of the art with
regards to providing reliability in distributed systems.
A BFT system implements a replicated state machine~\cite{schneider-computingSurv-1990} 
typically consisting of $n=3f+1$ replica servers
that each provide a finite state machine and  execute operations from clients
in the same order. BFT systems assume a pessimistic failure
model, based on the classic Byzantine generals' problem~\cite{lamport-tpls-1982} which provides
agreement amongst a set of nodes where at most $f$ nodes display arbitrarily
incorrect behaviors, known as Byzantine faults.

BFT systems are attractive because they provide guaranteed safety and liveness
properties when the assumption of up to $f$ faulty nodes hold.  Early work on BFT systems was
widely considered to be impractical for use by real systems because they were either too slow to 
be used in practice or assumed synchronous environments that rely on known message
delay bounds.  However, the seminal work of Castro and Liskov~\cite{castro-osdi-1999}, 
published in 1999, 
changed this view.  This work proposed and implemented \emph{Practical Byzantine
Fault Tolerance} achieving impressive peak throughput of several tens of thousands (null)
operations per second, previously thought unattainable.  As has been noted by 
others~\cite{clement-nsdi-2009}, over the last twelve years, 
the research community has seen a flurry
of excitement with several efforts to improve the performance and/or cost of BFT replication
systems.  These efforts include studies aimed at increasing throughput or reducing latency of client 
requests~\cite{yin-sosp-2003,kotla-dsn-2004,abd-el-malek-sosp-2005,cowling-osdi-2006,distler-eurosys-2011,garcia-eurosys-2011,kotla-sosp-2007,vandiver-sosp-2007,clement-nsdi-2009,wood-eurosys-2011},  
efforts to reduce the number of replica servers needed to withstand $f$ faults to achieve
lower replication cost~\cite{yin-sosp-2003,distler-ndss-2011,wood-eurosys-2011},  
and efforts to boost the robustness of the protocol under both faulty servers
and faulty clients~\cite{amir-dsn-2008,clement-nsdi-2009}.
A majority of these systems~\cite{yin-sosp-2003,kotla-dsn-2004,kotla-sosp-2007,amir-dsn-2008,clement-nsdi-2009,garcia-eurosys-2011,wood-eurosys-2011}
are direct dependents of the Castro and Liskov system, hereonin referred to as the
PBFT approach
(for Practical Byzantine Fault Tolerance).  Both the implementations and evaluations of
these systems depend on the initial PBFT code base. 

Despite PBFT's attractive correctness guarantees, BFTs are still rarely used in
practice.  This is unfortunate, given the ever-increasing need for reliability
in real-world distributed systems.  More and more applications  
require utmost security and reliability
to be both trustworthy to users and successful in use (e.g, electronic voting and 
digital preservation). 
The lack of wide deployment  of state-of-the-art BFT technologies is also puzzling.  The open-source 
PBFT code initially provided by Castro and later modified by others has been publicly 
available, improved, and fine-tuned for several years, 
and while readily
sized up by the academic community for research purposes, it has not been used in practice
in real-world systems.

In this paper, we examine, from the perspective of an application developer, 
the \emph{practicality}, i.e., feasibility, of using the PBFT protocol and  
accompanying implementation  
to provide provable reliability guarantees for a
real-world application.  Our motivating application is a state-of-the art electronic voting system,
offered as a public Internet service.  The current version is centralized~\cite{kiayias-acsac-2006}.  Given the
critical nature of the application, our aim is to build a system that has \emph{no centralized
component}.  Every aspect of the system's design should be distributed to avoid single
points of attack and failure.   
Our aim is to leverage the correctness guarantees provided by PBFT systems to improve
the security and reliability properties of the system.  
In such a system, clients (on behalf of users/voters) connect to the voting service,
view the election procedures to which they have a right to participate, send the user's vote,
and potentially reconnect at a later point to view the progress and/or results of the election.
Our aim has been to gauge, from the perspective
of a developer in need of providing reliability beyond simple crash-fault recovery, 
how easily the PBFT approach and accompanying system could be molded to fit the 
application developer's needs.

We have focused on the original PBFT implementation for several reasons.  First, 
over the past twelve years,
the majority of research efforts on improving BFT systems have relied on 
the PBFT approach and implementation.  This is the most stable code base that is 
publicly available and has been fine-tuned and improved 
over several years and by several developers.  
Second, even as the debate over improving BFT systems
continues, the interface to application developers provided by the PBFT middleware remains
the same. This means that 
any later developments in the PBFT system suite can be easily leveraged by applications.  Since
later systems are not as fine-tuned as the original PBFT code base, we have chosen the original
for more stability.  Third, our particular electronic voting application is written in C;  the 
PBFT code base is written in C++.   A recent effort, called UpRight~\cite{clement-sosp-2009}  
aimed at easing 
the application developer's effort to make use of BFT technology is written in Java,
still has several key features missing (e.g., view changes are unimplemented), and seems
to be a work-in-progress that has not seen much development in the last year and a half.  Thus,
for a developer wanting to leverage the attractive reliability guarantees of BFT \emph{now},
the original PBFT system offers the most promise.

We describe our experience trying to leverage the PBFT approach and code base to
enhance the reliability of our evoting application.  We describe several obstacles we 
encountered and drawbacks we identified in the PBFT approach.  
One key drawback we identified is that PBFT-based systems assume static membership --
ie., clients and replica servers know each other apriori before system initialization.  
Most Internet services
require support for dynamic client management, particularly when the number of 
envisioned clients is large.  The PBFT literature (original as well as all subsequent
descendants of PBFT) does not address this issue.  
Another key drawback is that PBFT leaves
state management completely to the application developer, who is required to manually manage
a raw memory region, while also issuing notifications to the library before changing memory contents.
This may be fine when developing system services, but is not a very convenient base for an application.
Additionally, PBFT treats a replica server's memory as stable storage, by assuming 
the use of uninterruptable power supplies  ~\cite{castro-osdi-1999}. 
Many Internet application services, particularly an electronic voting system, cannot afford
to rely on this assumption and instead require traditional ACID semantics to ensure
data stored is consistent and persists despite crashes and faults.  The PBFT system suite
leaves state management to the application developer.   This means that an application developer
wishing to make use of an available legacy database to provide the required ACID semantics is faced
with the decision of implementing from scratch these semantics into the application 
or retrofitting the BFT middleware
to interface with and support the legacy database.

In addition to the above, we describe a number of other drawbacks including: the mechanism
used by PBFT to handle nondeterminism in applications, the lack of support for stronger
cryptography, the lack of support for web-based applications, and others.  The description
of our experience may seem pedantic, with many minute low-level details, but we provide these
here to give the reader a clear understanding, from a holistic systems perspective, of the 
obstacles faced by a developer trying to put the PBFT system to real, practical use.
These are details that are often considered ``not important enough'' to warrant attention
and space in many research papers (and prototype implementations, for that matter), 
usually due to time and space constraints. Nonetheless they can trip up a third-party
developer hoping to make use of the novel research prototype.   In practice, it
is the details that make or break the widespread deployment and use of a system.

We find that while many of the obstacles we describe could be overcome with a ``better''
or ``revised'' BFT middleware 
implementation that is tuned specifically for the needs of the particular application, they 
require significant engineering effort and time.  Even less encouraging is the fact that
the performance implications
of the changes required to meet the application's needs are unclear.  For example,
we describe how we overcome the first two drawbacks above.  
While adding support for dynamic
client management does not significantly affect system performance, measured in \emph{null} operations per second,
retrofitting the
PBFT middleware to support a legacy database 
reveals a throughput performance of \emph{real} operations that is two
orders of magnitude smaller than the  \emph{null} ones, advertised
by prior BFT studies. 

To date, only two 
publications on BFT that we are aware of have noted that reporting null operations per second
as throughput is not representative of real applications and thus not helpful to the end-developer~\cite{sen-nsdi-2010,wood-eurosys-2011}.  This is understandable, as the focus
of most BFT research efforts has not been on end-application use but on improving the BFT 
middleware itself.  Nonetheless,  a developer faced with having to
make a slew of modifications to the BFT middleware to get an end-system that has unknown
performance properties is hesitant to invest the effort to do so.  

This paper makes the following contributions:

\begin{itemize}
\item We identify a number of drawbacks in the PBFT protocol suite, from the
perspective of an end-application developer trying to leverage PBFT reliability 
guarantees and we describe a number of potential solutions to address these.
The sheer number of drawbacks severely affects the ease with which a developer
can leverage the PBFT approach.

\item We present changes we made to the PBFT protocol and implementation
to enable dynamic client management, a must for many Internet service applications
in use today.  We show that these changes can be made with minimal additions to the
PBFT protocol, thus not affecting its provable reliability guarantees.  We demonstrate, via
empirical experiments, that support for dynamic client management can be achieved
with minimal performance impact.

\item  We evaluate the performance impact of retrofiting the PBFT middleware to 
support ACID semantics via a widely-used legacy database to ease the state management burden
of many applications requiring these semantics.  
We evaluate the impact on performance of this change, and show that for \emph{non-null} operations,
the throughput can be many times smaller than the tens of thousands of \emph{null} operations
per second presented in prior PBFT-based studies.  

\end{itemize}


\section{Background}
\label{sect:backg}

\subsection{Original algorithm}
The Castro-Liskov algorithm for Practical Byzantine Fault Tolerance \cite{castro-osdi-1999} 
(abbreviated as \emph{PBFT}) is a replication algorithm that can tolerate arbitrary 
faults. It is based on State Machine Replication \cite{Lamport78,schneider-computingSurv-1990}
where transitions are applied to an instance of the application's state and result in a new, 
deterministic instance of the state. The general idea is that a group of replicas form a 
static group that provides a service. At each instance in time, one of them is the primary and is responsible for 
sequencing the requests, providing total order. This in turn guarantees linearizability~\cite{TOPLAS:HerlihyW1990}, 
which is a correctness condition for concurrent objects where a concurrent computation is equivalent 
to a legal sequential computation.
A view is the epoch where the primary is stable. The remaining replicas monitor client requests and the primary's behavior and, if the latter is found misbehaving, begin a view change procedure and elect the new primary.

The algorithm is asynchronous and provides liveness and safety guarantees when less than 
a third of the replicas are faulty. More specifically, to tolerate $f$ Byzantine faults, the group needs at least ${3f+1}$ members. Safety, formally proved by using the I/O Automaton model \cite{Lynch:1996:DA}, guarantees that replies will be correct according to linearizability. Liveness assures that clients will eventually receive replies to their requests. The algorithm does not rely on synchrony to provide safety but does rely on a weak synchrony assumption to provide liveness: that \emph{delay(t)} does not grow faster than \emph{t} indefinitely. Here, \emph{delay(t)} represents the time interval between initial message transmission \emph{(t)} and message delivery to the replica process. For the protocol to be live, the client is expected to keep retransmitting its request until it finally obtains the reply. Further assumptions include independence of node failures and inability of an attacker to subvert cryptographic protocols.

In normal operation, the client sends a request to the primary. The primary assigns a monotonically increasing sequence number to the request and begins a 3-phase agreement protocol with the other replicas, at the end of which each node executes the request and directly transmits the reply to the client. The latter will accept the reply as correct only when $f+1$ replies much. The 3-phase protocol consists of the exchange of the following messages, where the target of a multicast is the set of replicas:
\begin{enumerate}
 \item \emph{Pre-prepare}, multicast from the primary, which assigns a sequence number to a request and forwards its contents
\item \emph{Prepare}, multicast by each replica, agreeing to the sequence number assignment 
\item \emph{Commit}, multicast by each replica, which helps guarantee total ordering across views
\end{enumerate}

After the commit, each replica will execute the request and transmit the reply directly to the client.
In all above message exchanges, the sender is expected to sign the contents with his private key.

\begin{figure}[h]
\centering
  \includegraphics[width=0.47\textwidth]{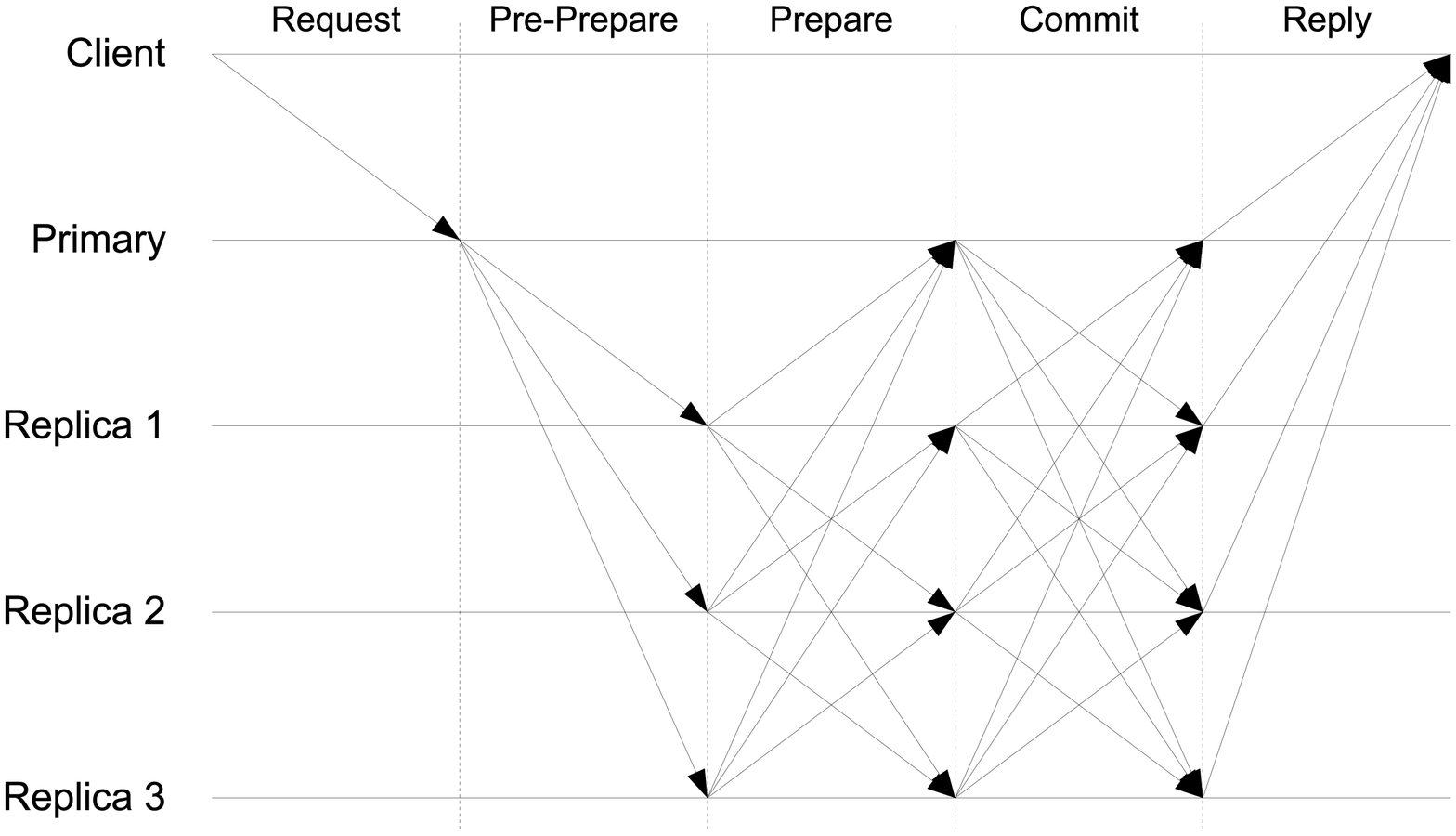}
\caption{Normal PBFT operation}
\label{figure:bftorg}
\end{figure}

This operation is depicted in Figure \ref{figure:bftorg}.

Certain optimizations were applied by Castro and Liskov to this basic mode of operation in order to 
improve the latency and throughput of the system. First of all, the use of 
asymmetric cryptography was reduced, by introducing Message Authentication Codes. 
The client assigns a different key to each replica and sends the key to it, 
signed with the node's public key. From then on, all requests are accompanied 
by an `authenticator', which is a structure that contains one MAC for each replica. 
This considerably boosted performance, as we be confirm in Section~\ref{sect:eval}.  
Another optimization is the tentative execution of requests before the commit phase. 
The client cooperates in this mode of operation as it expects 2f+1 tentative replies 
(marked as such by each replica) instead of the normal f+1. If such a quorum is not 
assembled, the client simply retransmits the request message. As the replicas will in turn 
retransmit the last reply for this client (which by now should be marked as stable, since
the Commit phase should be over), a smaller quorum of f+1 stable (non-tentative) replies may be enough.

Yet another optimization is the special treatment of \emph{read-only} and \emph{big} 
requests. A request is considered \emph{big} if its size exceeds a configurable 
threshold, while the \emph{read-only} status is explicitly set by the client. These 
differentiated requests are multicast from the client to all replicas, to relieve 
this burden from the primary. This mechanism is utilized by default to the maximum extent,
by defining the threshold to 0, resulting in all requests treated as \emph{big}.
The \emph{read-only} requests are specially treated 
and are executed as soon as they are received, sequencing permitting, of course.
Finally, \emph{request batching} is employed to minimize network usage and agreement 
latency. A \emph{congestion window} is defined as the number of requests that have 
been received but not yet executed by the primary; its size is an adjustable parameter 
of the system. When the primary receives a request message, it calculates the difference 
between the last locally executed sequence number and the sequence number assigned to the new 
request. If this difference exceeds the defined \emph{congestion window}, it postpones 
issuing the \emph{pre-prepare} message, giving itself time to catch up on request execution. 
Once it does, it includes in a single \emph{pre-prepare} message, as many outstanding request 
messages as possible, thus minimizing latency due to individual agreement. 
Note that batched requests capture parallelism from different clients, as each client 
is allowed a single outstanding request only.

An implementation of the protocol was developed by the author, Miguel Castro, and published as 
open-source along with his dissertation. The environment chosen was:
\begin{itemize}
\item Linux as the platform (but mostly POSIX compliant)
\item C++ as the base language
\item UDP as the network protocol
\item An implementation of the Rabin cryptosystem for asymmetric cryptography
\item An implementation of UMAC32 for MAC operations
\item An implementation of MD5 for digests
\end{itemize}

This implementation defines application ``state'' as a single continuous virtual 
memory region. In fact, it splits this region in two, the first part for the 
internal library needs and the remaining for the application. The library 
has a subsystem that manages the synchronization and checkpointing of this 
state using copy-on-write techniques and Merkle (hash) trees \cite{merkle87}. 
The general idea is that the state is divided in pages of equal length.
A hash tree is formed where the leaves are the actual 
data pages while the inner nodes are the hashes of their children (either of the 
data pages at level \emph{height-1}, or of the hash text at smaller depths). 
At the root, a single digest uniquely identifies the complete memory region. 
A checkpoint message communicates this root hash to the rest of the replicas, 
to agree that the state is properly synchronized. If a peer finds itself 
out of sync, an efficient tree walking algorithm is started from the root, 
to identify the (hopefully few) data pages that are different and have 
them retransmitted by the rest of the group. 


The server part of an application wishing to use PBFT services, is 
expected to initialize the library and then wait for up-calls from it, 
to service requests and produce replies. While executing, it has free 
read access to arbitrary memory regions inside the ``state'' managed 
by PBFT, but is expected to notify the library {\bf{before}} making any changes.

\subsection{Reasoning about the default implementation}

It is very hard to reason about the behavior of a distributed system 
when it is run on multiple hosts, without a common clock. Although 
solutions such as vector clocks exist, it would 
be too intrusive to retrofit them in the existing library, just for the sake 
of monitoring its operation. To this end, we modified 
the library to be able to run multiple times on the same host, using different 
port numbers. We also created a log of all messages exchanged 
between replicas that, given the common clock, allowed us to reason about 
the behavior of the system. All further observations are based on this groundwork.

\subsection{Authenticators and Erratic Recovery Behavior}
In an attempt to closely monitor and better understand the recovery process, 
we stopped and restarted a replica, using the default optimal configuration. 
We immediately witnessed erratic behavior in the recovery process, which 
started and re-synchronized the state to the latest checkpoint, but was unable 
to execute the few requests remaining in the log after that point, because 
they failed the authentication test. What we found after further 
investigation was that the use of authenticators, introduced for 
efficiency, impeded the recovery process, because the transient state of 
the restarted replica had no recording of the authenticators to use for 
validating client requests. The solution the existing system implements, 
is the blind retransmission of the authenticators from each node to all 
replicas, based on a timer. This way, once the recovering replica 
receives the authenticators of the clients, it will be able to resume the recovery 
process from the next checkpoint. The only way to lower the time frame for 
this service interruption, is to reduce the authenticator retransmission 
timeout, which results in increased load for the network. We investigated other solutions including
on-demand retransmission of the authenticators;  we did not pursue this however, 
because retransmissions could introduce denial-of-service vulnerabilities, as a faulty replica could simply 
bombard the clients with authenticator retransmission requests.   

\subsection{PBFT Behavior on UDP Packet Loss}
The definition of a Byzantine fault, for the PBFT library is any possible fault, 
including an error as trivial as a UDP packet loss.  This creates interesting behaviors. 
We observed that UDP packets were indeed lost in our experiments, even in the loop-back 
interface, due to congestion caused by stress-testing the system. The impact of this is 
profound, as such an error will leave a replica lagging behind in transaction execution 
and will cause the recovery process to commence on the next checkpoint. Unfortunately, 
although this apporach is theoritically very elegant, it is unacceptable for a production
environment to lose nodes from such trivial errors.

One of the optimizations described above, regarding the special handling of 
\emph{big} requests, combined with a trivial UDP packet loss, can greatly 
affect the robustness of the system. In this case, \emph{big} requests are multicast 
to all replicas only once, from the client. The primary will then use only the 
digest of the request body for further communication with the rest of the 
replicas. Consider what happens if one of the packets traveling from the client to one 
of the replicas is dropped on the way. All replicas will begin the three-phase 
protocol to commit and execute the request, but when execution time comes, 
the replica that missed the request body will be unable to execute, and 
will be stuck at this point until the next checkpoint arrives and the recovery 
process kicks in. For a request not marked as \emph{big} though, the 
process is different and more stable. Here, if the request from the client 
to the primary is dropped, the client will timeout and retransmit the request, 
resulting in a request execution workflow where either all or no replica at all 
participates. Even in this case, a replica-to-replica packet loss would 
again result in interruption of service for one of the replicas, but perhaps
in some environments, one can assume this to be less frequent than 
client-to-replica packet loss.

\subsection{PBFT Handling of Non-determinism}
In the original PBFT implementation, a feature was introduced to resolve the 
non-deterministic characteristics of most applications. The primary makes 
an application-specific up-call, which returns a set of values that are 
attached by the primary to the Pre-Prepare message. This data becomes common to 
all replicas executing the request, thus providing deterministic behavior on 
request execution. Subsequent work on the PBFT protocol ~\cite{castro03base} added an extra 
mechanism to validate this data on each replica. A new application-specific 
up-call was established that, when passed the non-deterministic data, is 
expected to validate it and return success or failure. The idea is, for example, 
that the primary attaches the system clock to the Pre-Prepare message, and 
each replica validates the passed value against its own clock to make sure it is appropriate.

However, the handling of non-determinism described above introduces a subtle issue.  
It is not always clear how the application can validate the non-deterministic data passed
to it via the new upcall.  
The hurdle for such a validation is the instance in time it is supposed to happen. 
In the normal, fault-free lifetime of a request, the validation happens as soon as 
the Pre-Prepare message is received, which is almost immediately after it is transmitted. 
Thus validating against a time delta is viable. However, when a request is replayed from 
the log during recovery, the time drift can be quite large and validating using a time delta
will fail and impede the recovery process. A solution to this issue would be to 
differentiate message processing for the recovery process and completely skip
non-deterministic data validation during recovery. This however is again a non-trivial
exercise, as message execution in PBFT is completely orthogonal to its origin.

\section{PBFT Deployment Drawbacks and Obstacles   }
\label{sect:revisedBFT}

In this section, we present in detail a number of the obstacles
we encountered in trying to leverage the PBFT approach and implementation
for our Internet evoting service.   While some of the details may seem
pedantic and low-level, we include them here to give the reader a clear idea of
the kinds of issues an application developer must face in porting his application
to a BFT version.

\subsection{Dynamic client membership}
\label{sec:dynclients}

The existing PBFT protocol and implementation assumes completely static membership 
where each node in the system, client or replica, needs a priori knowledge of the 
address, port, and public key for every other node.  For many applications, particularly Internet
service applications with a large number of clients, such a closed
system does not suffice.  Our goal is to remedy this to enable clients to join and leave the 
replicated service dynamically, while letting the replicas remain statically bound to one another. 
The end result is that clients only need information regarding replicas, but no 
information regarding other clients, allowing for a more scalable deployment.

To achieve support for dynamic client membership, the replicas need to identify each 
client in an identical (deterministic) manner.  This leads us to store the client 
identifiers in the shared state of the service (i.e., in the continuous memory region). 
When a client requests to join or leave the group, each replica needs to 
process the request using the same version of the shared state. Thus, all such client 
requests need to be totally ordered, at least with respect to one another.

We define two special system requests, namely a \emph{Join} and a \emph{Leave}, which 
follow the same life-cycle as all other application-level (client) requests. This 
results in a single total order across all requests, application or system, fulfilling
our requirement. 
The \emph{Join} and \emph{Leave} system requests are processed by the 
middleware library and are invisible to the application.

We introduce a level of indirection between what the PBFT library already uses as 
a node identifier and what the client reception module assigns to new clients, 
for efficiency of message evaluation. Instead of using a single address range of 
[$0$..$max\_clients$], an arbitrary identifier is assigned to each new client and 
a table maps this number to the index in the array of client and server node entries. 
This way, when a  client request arrives, the system first checks to see if the 
identifier exists in the redirection table before going into the more lengthy 
process of verifying its signature or authenticator.

Originally, our idea was for the client to multicast a simple Join system request 
to all replicas, carrying its address, its public key and a random nonce, signed 
with its private key. Each replica would assign the same new identifier and 
transmit it back in the reply. However, nothing stops a malicious client from 
initiating an infinite number of connections, using phony addresses, thus 
exhausting the bounded maximum number of node entries in each replica. To address 
this vulnerability, we improve the connection process by splitting the Join 
operation into two phases. In the first phase, the client submits its data as previously 
described and awaits a challenge.  Upon receiving the challenge, the client calculates 
a response and transmits it back to the replicated service in the second phase of 
the Join. Only then will the replica add the client to the system as a full member. 
This approach ensures the client indeed owns the address he claims, as receiving the challenge
is imperative to compute the response.

We also add an application-level identification buffer to the Join message. 
This buffer is passed to the application for authorization.  It might include, for example,
an encrypted user id and password.  The application then returns an identifier to 
be associated with this client (such as the user id). The middleware library will 
then guarantee that only a single session can be active at a time for this 
specific identifier, by terminating all previous sessions when a new one is 
established. This way, even in a distributed denial of service attack, the 
attacker can only establish as many sessions as the number of credentials 
he has managed to obtain.

The Leave system request is much simpler as it simply instructs each replica to 
remove the client from its internal tables. All further communication with 
the service is prohibited for this client.

We need timeouts to enforce cleanup of stale sessions once the node 
structures are full. To achieve some common ground regarding time across all 
replicas, all requests are timestamped with the time of the primary; when each 
request is executed, its timestamp is recorded for each client. When a join 
request arrives that cannot be serviced because the client/server node table is full, 
a cleanup process is started that will locate all clients with a last executed request 
older than the current join request minus a configurable threshold. All such sessions 
are cleared to make room for the new connection. If no such stale sessions are found, 
the new Join request is denied.

\begin{figure}[h]
\centering
  \includegraphics[width=0.47\textwidth]{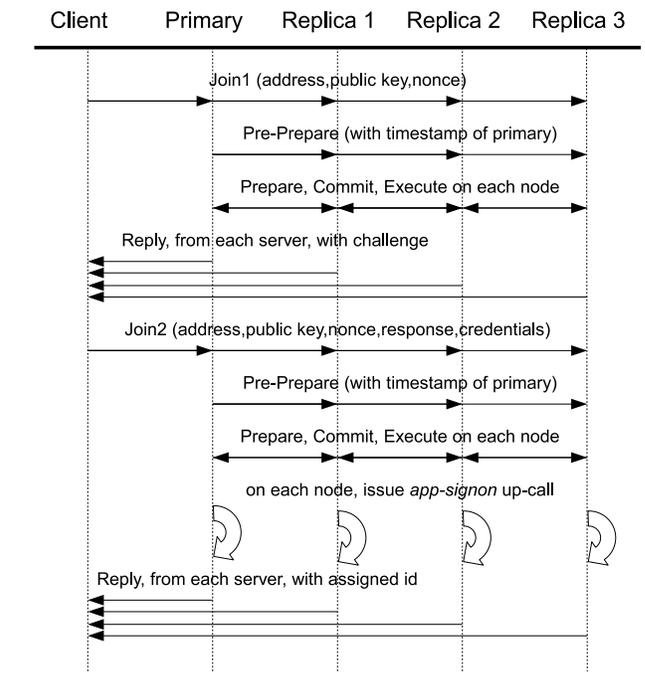}
\caption{PBFT dynamic client join sequence diagram}
\label{figure:bftjoin}
\end{figure}

The Join process is depicted in the UML sequence diagram shown in Figure~\ref{figure:bftjoin}.

Note we have enhanced the PBFT protocol with 
support for dynamic client membership without changing the inherent properties
and message exchanges of the protocol.  Thus, our changes do not affect the 
safety and liveness guarantees
offered by PBFT.

\subsection{A higher level state abstraction}
\label{sec:state}

In a replicated state machine, the term `state' is an abstract definition of the 
persistent workspace of the application. PBFT defines state to be a continuous 
virtual memory region where both the application and the middleware library store 
their non-transient state, in contiguous non-overlapping partitions. The middleware 
library has full access to this memory region while the application code is not executing, 
since it is responsible for managing replication and synchronization of this state 
across replicas. The application, on the other hand, has free read access to it, 
but is required to notify the library before making changes to any region, 
thus permitting copy-on-write optimizations of state synchronization.

While the above approach relieves the application considerably from having to deal with 
state synchronization, it creates a number of questions which the application developer must
face:  What can a modern application do with just a pointer to a memory region? 
How is this state persistently stored on disk when the service stops? And how does the developer avoid
the havoc caused by a misbehaving application which fails to notify the library before modifying memory? 

To answer these questions in a satisfactory manner, we decided to adapt an embedded 
relational database engine, to intervene between the PBFT middleware library and the 
application. This way, the application will have SQL-level access to its state and 
the embedded engine will take care of interfacing with the PBFT library to satisfy 
its requirements. 

In our search for an embedded relational database engine, the major feature we were after 
was storage of data in a single file, which we could map to virtual memory. 
We selected SQLite \cite{SQLite/site} because it exhibits this feature and because it is mature and
widely deployed.  SQLite is an embedded, in-process library that implements a self-contained 
relational database engine using SQL as its command language and a C call level 
interface for the application. It stores all data objects in a single database file 
that is binary compatible across machine architectures (endianness) and word sizes.

\begin{figure}[h]
\centering
  \includegraphics[width=0.47\textwidth]{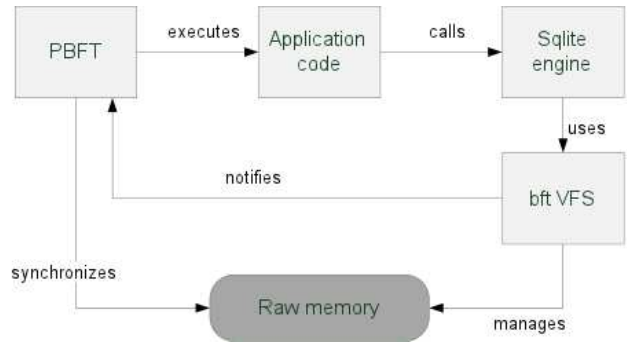}
\caption{SQLite with its VFS inside a PBFT application}
\label{figure:sqlvfs}
\end{figure}

In SQLite's quest to be a multi-platform product, the authors have defined an abstraction 
layer called VFS (Virtual File System), that sits between the relational engine and the 
operating system. By hooking into this subsystem, we not 
only can manage memory mapping and perform PBFT-required memory modification notifications, 
but also re-implement non-deterministic functions, such as system time and random values, 
by using the upcalls described in Section~\ref{sect:backg}.  
Interaction with VFS is illustrated in Figure~\ref{figure:sqlvfs}.


SQLite uses two disk files to manage the database, for reliability reasons. The 
first file is the actual database, which we map to virtual memory. The second file
is the rollback journal (or write-ahead-log, in a different mode of operation), 
which is used to rollback failed transactions. We left this second file to be stored 
on disk, since it allows the engine to recover in the case of system failure and it is 
not actually part of the application state. In any case, the database file is 
synchronized with its disk image on transaction commit.

We gain many advantages with this approach.  First, a committed transaction will be 
durable, even in the case of a system crash. That is, when the replica node 
restarts operation, its state will include the last committed transaction, and 
PBFT recovery will commence from this point. Second, even if the node is to be removed 
from the replicated service, its data will be usable on its own, being just another 
database file. Moreover, an uncommitted transaction will be rolled 
back on the next attempt to access the database file, from the replicated 
service or on its own. These advantages are simply the by-product of the ACID 
semantics that SQLite provides and excellent reasons why developers will likely
want to take advantage of it.  

One obstacle we faced was that, while SQLite can freely manage the growth and 
shrinkage of its database file, PBFT is not so permitting, because it requires 
knowledge of the size of the memory region that represents the state, during its 
initialization. To alleviate this, we use a sparse file that is defined to be 
a large enough size on initialization, without actually occupying that space 
on disk, a solution that is reasonable in modern 64-bit operating systems 
with large virtual memory address ranges. 

The application code now simply passes the name of the database file to the PBFT 
initialization function responsible for starting up the replica server and setting up any
data structures needed by the middleware.  The function returns to the application code
a standard SQLite database handle. Using this handle, the application can call 
standard SQLite library functions (e.g. sqlite3\_exec, sqlite3\_prepare\_v2, 
sqlite3\_step) to access the database while executing during the appropriate PBFT 
upcall. This way, an application already using SQLite is immediately portable 
to the PBFT middleware with only  minor changes to the initialization code.

\subsection{Remaining issues}
We now describe a number of remaining issues we encountered in the 
process of applying the PBFT approach to our electronic voting application
service.

\subsubsection{Cryptography}

Applications requiring strong cryptography, such as private key generation and 
storage on the server side of the application, are not well supported by 
the current PBFT implementation. For key generation, strong random 
values are required. Unfortunately, even if the primary obtains such strong randomness 
from its local OS services, for example via /dev/random, there is no way such 
values can be verified from the remaining replicas, by their very definition 
of being random. Because of this, an adversary can obtain access to one of 
the execution replicas, wait until it becomes the primary and use 
predetermined values instead of random values. In this manner, the adversary can
trigger the generation of well-known 
private and public keys and thus violate 
confidentiality. To alleviate such attacks, one solution would be to enforce a 
threshold signature scheme ~\cite{DesFra89} for such 
authentication
requirements, provided for by the middleware library.
In such a scheme, private key information for each replica would never be transmitted over the
network, as it would not be stored in shared state. In a $(f+1, n)$ (where $n=3f+1$) threshold signature 
scheme, the set of $n$ replicas would collectively generate a digital signature despite up to $f$ byzantine faults.
Of course, the PBFT protocol would have to be modified to provide for such cryptographic operations.

Another confidentiality issue is the matter of protecting storage of sensitive 
information. This has been studied by Yin et al~\cite{yin-sosp-2003}, who propose separating
the agreement part of the PBFT protocol from the execution part, while also adding an intermediate
cluster of `privacy firewall' nodes. In this layout, $3f+1$ \emph{agreement} nodes receive the 
client requests and forward them to $2f+1$ \emph{execution} nodes for execution. To ensure that a
faulty execution node cannot disclose sensitive information, an $h+1$ rows by $h+1$ columns \emph{privacy firewall} 
set of nodes is positioned between the agreement and execution cluster, which allows tolerating up to $h$ 
faulty firewall nodes. This obviously increases both deployment complexity and request execution latency.


\subsubsection{Stateless applications only}
The current implementation of the PBFT protocol purposely ignores the notion of 
client-specific state. This, however, severely limits the target applications to 
those that are either stateless by nature, or manage session state on their own 
using their global state abstraction; the latter will need to pass session 
identifiers inside the request and reply bodies, without any assistance from 
the middleware library. This is not an inherent limitation of the State Machine 
Replication approach.  It is simply a consequence of the lack of appropriate 
mechanisms in the PBFT library. With our addition of application level sign-on 
messages to the protocol, resulting in identification of specific sessions, 
a library-level subsystem can be developed that will map parts of the state 
to a specific session. This would enable easier porting of stateful 
applications to the BFT world.

\subsubsection{Web applications}
Our end goal is to provide a web application to end users, which provides them 
hassle-free access to the server counterpart of the evoting service.  We aim to achieve this 
without sacrificing BFT semantics. To this end, the browser-hosted part of the application, 
typically written in JavaScript, will have to directly access each and every replica. 
This communication however cannot be carried over UDP because this protocol is not 
allowed in the JavaScript runtime environment. Moreover, binary messages are highly 
inconvenient in this context. Higher level protocols, such as WebSocket, and 
structures like JSON or XML need to be used. Support for these technologies 
needs to be incorporated in the middleware library, a task not so trivial because 
of the need to switch from a point-to-point message-based communication to a connected 
channel-oriented communication. Additionally, cryptographic functions will need to 
be available in the browser-hosted client part, which requires transitioning from 
Rabin to more widely available cryptosystems, such as RSA.

Additionally, we aim to have the replicas located in different physical locations, to obtain
real independance of faults caused by network partitions. This requirement dictates operation in a
Wide Area Network environment, where the quadratic message complexity of PBFT will most probably
prove costly regarding request latency. Although we tried to simulate a WAN deployment scenario
using BFTsim ~\cite{conf/nsdi/SinghDMDR08}, the simulator could not scale to a large enough number of nodes
($> 100$) to obtain meaningful results. This issue is already studied in ~\cite{Amir05steward:scaling}, 
though no open source implementation is readily available.

{\bf Summary:}  The above issues can be overcome, but require a significant amount of engineering
effort.  An application developer wanting to leverage and deploy PBFT \emph{now} is likely
to be unwilling to invest the time and effort required to retrofit the PBFT approach to match
the needs of his/her application.     

\section{Evaluation}
\label{sect:eval}

In this section we present empirical measurements of the PBFT library, both with and without
our modifications supporting dynamic client and seamless state management for applications
requiring ACID semantics provided by a legacy database.

We test the PBFT library and our modifications to it on a cluster of 8 machines
connected with a 1GB Ethernet switch. The first four machines are Intel Xeon E5620 
at 2.40 Ghz under CentOS 5.5 with Linux kernel 2.6.18-194. The remaining four are 
Intel Core 2 Duo E6600 at 2.40 GHz under Debian 5.0 with Linux kernel 2.6.26. All eight 
machines run 64 bit versions of their corresponding operating systems.
Ping roundtrip time is measured at 134-183 nanoseconds between all hosts. Bandwidth 
is measured, using iperf, at 938 Mbits/sec.
For all tests, we generate a server and client executable using a particular library configuration
set so as to measure the effect of turning on or off a particular optimization and/or modification.
We designed the client to connect to the library and wait for a signal. On signal 
reception, it records the current time, starts its operation and then measures and reports elapsed time.
To coordinate all processes running on different hosts while at the same time collecting and
aggregating measurements, we implemented a test framework using Python and netcat, where the 
latter runs on each host and allows a single controller to submit scripts (i.e., experiments) 
and collect the results.

\subsection{Non-SQL Experiments}

We first conduct an experiment without the SQL state abstraction modifications we made
in order to benchmark 
the plain PBFT implementation. Our goal is to measure the impact on system throughput of 
turning on/off the optimizations
described in Section~\ref{sect:backg}.    Recall that the use of certain optimizations 
(such as the use of MACs and special handling of \emph{big} requests) increases
performance at the cost of  
decreased robustness (e.g., slow recovery) of the system.

\begin{center}
\begin{table*}[ht]
\small
\hfill{}
\begin{tabular}{l c c c c r r}
\hline
\textbf{Name} & \textbf{Static client mgmt} & \textbf{Using MACs} & \textbf{All requests treated as big} & \textbf{Batching} & \textbf{TPS} & \textbf{StDev}\\
\hline
sta\_mac\_allbig\_batch & Yes & Yes & Yes & Yes & 17.014 & 66 \\
sta\_mac\_allbig\_nobatch & Yes & Yes & Yes & No & 1.051 & 56 \\
sta\_mac\_noallbig\_batch & Yes & Yes & No & Yes & 3.030 & 57 \\
sta\_mac\_noallbig\_nobatch & Yes & Yes & No & No & 1.109 & 103 \\
sta\_nomac\_allbig\_batch & Yes & No & Yes & Yes & 1.291 & 4 \\
sta\_nomac\_allbig\_nobatch & Yes & No & Yes & No & 1.199 & 12 \\
sta\_nomac\_noallbig\_batch & Yes & No & No & Yes & 992 & 2 \\
sta\_nomac\_noallbig\_nobatch & Yes & No & No & No & 1.186 & 7 \\
nosta\_nomac\_noallbig\_batch & No & No & No & Yes & 988 & 1 \\ 
nosta\_nomac\_noallbig\_nobatch & No & No & No & No & 1.205 & 1 \\
\end{tabular}
\hfill{}
\caption{PBFT library configurations we test. TPS is transactions per second, where a transaction is simply
a null request.  Null request and null response sizes are 1024 bytes.}
\label{table:libbyzConfigurations}
\end{table*} 
\end{center}

We generate and test a series of PBFT library configurations, shown in 
Table~\ref{table:libbyzConfigurations}.  The first configuration is the
default configuration preferred and recommended by Castro, with all optimizations enabled, including
the use of MACs, special treatment of all requests as \emph{big} requests, and request batching. 
Since batching is the only optimization for which we did not observe faulty behavior, we isolate
it and test all other combinations of configurations with batching enabled and disabled, to show its
impact. The last four rows of Table~\ref{table:libbyzConfigurations} depict
the most robust configurations (use of MACs and big request handling turned off).  Since our particular
application has stringent security and reliability requirements, we choose to measure the impact
of adding support for dynamic client management using these configurations.  We believe other 
Internet service applications
with similar high security and robustness needs would need to run the PBFT library using these 
configurations.
The client and server programs built to measure throughput transmit null requests and responses  
of varying sizes, of 256, 1024, 2048 and 4096 bytes. 
We test the system using 12 clients spread evenly across 4 machines while being serviced by 
4 replicas, each running alone on a single host. 
In all cases, IP-level multicasting was turned off, as the networks we are targeting (WANs) do not support it.

\begin{figure}[h]
\centering
  \includegraphics[width=0.47\textwidth]{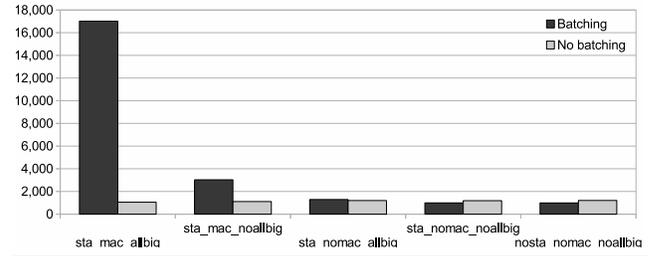}
\caption{PBFT tests}
\label{figure:bfttests}
\end{figure}

The results for varying request and response sizes are similar, so for brevity we show a representative
plot, for size of 1024 bytes in Figure~\ref{figure:bfttests}.

From Table~\ref{table:libbyzConfigurations} and Figure~\ref{figure:bfttests}, it is clear that
the first configuration, which is the default configuration of the PBFT library with all optimizations
turned on achieves the best throughput performance.  In our experiments, this configuration achieves
approximately 17000 null operations per second, while for the most robust configurations the throughput
drops to about 1000 null operations per second.

We observe that disabling the batching optimization seriously affects performance when 
using MACs. When switching to signing with private keys, the delay introduced is so large
that batching can no longer assist in any way.
Moreover, when disabling big request handling, performance drops to 18\% of the optimal, 
while disabling the use of MACs causes performance to drop to 7.5\%  of the optimal respectively.
Disabling both big request handling and MAC use causes performance to drop to 6\% of the optimal. 
While we observe a difference in performance amongst these configurations where some subset of
optimizations is turned off, the bottom line is that performance takes a big hit when turning off 
any of the optimization.    However, for an application with high security requirements, we conjecture
robustness is favored over performance.

We evaluate the impact on performance of adding support for dynamic client management 
using the most robust configurations.  
The performance decrease is 0,5\% (988 vs 992), which is negligible. 
This negligible decrease in performance is attributable to the cost of accessing the 
redirection table that converts assigned customer ids to indexes in the tables tracking participating
nodes (clients and servers).
We emphasize that the above tests are \emph{artificial} because they are testing ``null'' 
operations. The software on the replica spends no time executing application code; it 
simply manages the network protocol.   The large majority of prior BFT studies present
throughput in terms of null operations per second.  This is understandable as the focus is on 
providing a baseline benchmark against which varying BFT protocols can be compared, but is
not helpful to the application developer who needs to understand how the system would behave using
real application requests.

\subsection{SQL state abstraction experiments}
\begin{figure}[h]
\centering
  \includegraphics[width=0.47\textwidth]{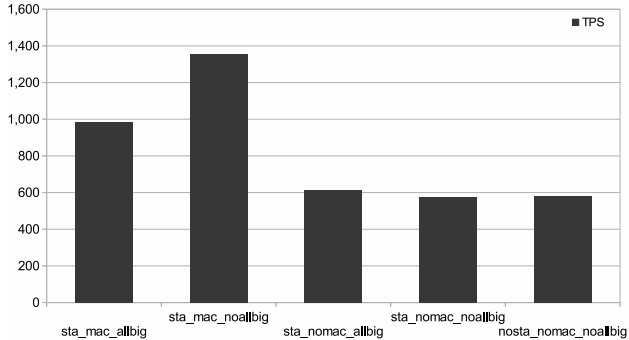}
\caption{PBFT + SQL benchmark}
\label{figure:bftsqlbenchmark}
\end{figure}

In this subsection, we evaluate the performance of adding seamless state management for applications
requiring ACID semantics provided by a legacy database.  Null operations are thus not realistic to 
use in this setting.   For our client application request we choose the insertion of a single row 
into a database table.  This is the operation our evoting service must perform to record a user's vote
in an ongoing election.    The tuple inserted into the database includes a simple key and value text
(representing voter identity and accompanying vote),
in addition to a timestamp and a random value.  We purposefully added the timestamp and random value to
test that replies are indeed identical across all replicas.
For this experiment, we enabled request batching and varied turning on and off the remaining options
(use of MACs, big request handling, and support for dynamic clients).  ACID semantics are provided
using the   
rollback journal mode of SQLite. Throughput performance, measured as database insertion transactions per second, 
is illustrated 
in Figure \ref{figure:bftsqlbenchmark}.
%

In this experiment, the big request handling optimization pays no dividends because the system now
spends time executing a real, non-null request which requires accessing the hard disk.  
This dominates the overall request execution lifetime.
At any rate, the most robust configuration with dynamic clients enabled is now at 43\% of the best (sta\_mac\_noallbig).
Since disk access is a big factor in this experiment, we perform two more experiments to isolate its impact.  
In these experiments, we measure the most robust configuration 
(where the use of MACs and big request handling are disabled) with dynamic clients and ACID semantics (as above) 
and we measure another configuration without ACID semantics (no rollback journal and no flushing to disk 
on each operation). The ACID version achieves 534 TPS while the No-ACID one scores 1155, an approximately 2x 
performance boost.



{\bf Summary:}  The optimizations turned on by default in the PBFT library, lead to
the high throughput numbers reported in prior studies, but as we have shown in 
Section~\ref{sect:backg}, using some simple fault scenarios (such as UDP packet loss),
the high performance numbers come at the cost of decreased robustness of the system.
Moreover, the performance numbers reported by a large majority of prior BFT studies
are based on a metric of null operations per second.   This is not a helpful metric
for the end-application developer, particular for a developer whose application makes
use of a legacy database for ACID semantics.

\section{Related Work}
\label{sect:related}

As cited in Section~\ref{sect:intro}, since the seminal 1999 publication on PBFT
by Castro and Liskov~\cite{castro-osdi-1999}, there has been a flurry of research activity
focused on improving the BFT middleware 
performance~\cite{yin-sosp-2003,kotla-dsn-2004,abd-el-malek-sosp-2005,cowling-osdi-2006,garcia-eurosys-2011,
kotla-sosp-2007,vandiver-sosp-2007,clement-nsdi-2009,distler-eurosys-2011,wood-eurosys-2011},
replication cost~\cite{yin-sosp-2003,distler-ndss-2011,wood-eurosys-2011},
and robustness under both faulty servers
and faulty clients~\cite{amir-dsn-2008,clement-nsdi-2009}.
A majority of these systems~\cite{yin-sosp-2003,kotla-dsn-2004,kotla-sosp-2007,amir-dsn-2008,clement-nsdi-2009,garcia-eurosys-2011,wood-eurosys-2011}
are direct dependents of the Castro and Liskov PBFT system.   Of all of these systems,
the only codebase that has been made widely available and refined for several years is
the PBFT system.  For this reason, we have focused on this system.  Since all PBFT 
descendants use the same codebase, the obstacles we encountered as application developers
in using the PBFT system apply to its descendents as well.  

We highlight, below, some related works that either directly focus on bringing BFT systems
closer to widespread deployment in real applications, or raise issues that affect the
practical deployment of our (and other) security-critical applications.




Wood et al.~\cite{wood-eurosys-2011} write ``no commercial data center uses BFT techniques
despite the wealth of research in this area'' and posit that this is due to the high cost
of replication required by BFT protocols.  They aptly point out that, for applications such as web servers
and database servers, it is the execution of client requests
and not the agreement of request ordering that dominates the performance of a BFT protocol.
They propose lowering the number of active
execution replicas to $f+1$ by using virtual machines as execution nodes and ZFS 
snapshots for quick state checkpointing. When the $f+1$ replicas produce 
inconsistent replies, a paused execution node is revived and starts executing 
requests immediately. The middleware library fetches the state needed by 
these requests on demand, to amortize the cost of state transfer. The paper claims
that for applications running over a WAN environment, the time to perform state transfer
is minimal compared to WAN latencies.  
The focus of the paper is on reducing replication cost while maintaining good performance.
While this is welcome for an application to be deployed in a data center, the paper does not address how 
the application developer can easily make use
of the system, stating simply that applications must be rewritten to take advantage of the 
system.


Clement et al.~\cite{clement-sosp-2009} introduce \emph{UpRight}, with the goal of
making it easy for application developers to convert a crash-fault tolerant application
into a BFT application.  
It includes a number of state-of-the-art BFT techniques, including separation of 
agreement from execution, insights from the Aardvark protocol~\cite{clement-nsdi-2009} 
on dealing with faulty clients and alleviating denial-of-service attacks, as well as 
more flexible state management (but not at such a high level as a relational 
engine). It also allows individual tailoring of crash-fault (Up) and 
arbitrary-fault (Right) tolerance. Unfortunately, it is still a work in progress.
with several key features missing (e.g., view changes are unimplemented) and 
does not seem to have seen much development since March 2010~\cite{upRight-project}, so it is not helpful
to a developer wishing to make use of BFT techniques now.

Several attempts have been made to address the inability of replicated 
BFT services to mesh with the rest of the infrastructure in today's multi-tier 
world. Merideth et al.~\cite{Merideth05} introduced \emph{Thema}, which aims to mask BFT 
complexity from the application developer of web services based applications. 
An agent, visible to the unaffected outside world, plays the role of the client 
of a BFT system. Additionally, a proxy collects the multiple out-call requests 
from the replicas of a BFT system, and issues the actual out-call on behalf of 
them, returning the reply when available. Unfortunately, both the agent and the 
proxy are centralized components which are inappropriate for applications such as ours
which require completely distributed design.

Pallemulle et al.~\cite{Pallemulle08} focuses on interoperability between BFT systems, 
while enforcing fault isolation and introduce a whole new protocol, named \emph{Perpetual} 
to achieve this.  Sen et al.~\cite{sen-nsdi-2010} in a system called Prophecy, designed  
to increase BFT performance,  introduce a \emph{Sketcher} component, 
that tries to trade space for performance, by storing a historical log of 
request/reply pairs and allowing the application to differentiate its requests, 
asking for possible log-based replies. In its distributed incarnation, 
\emph{D-Prophecy} is simply an attempt to avoid re-execution of repetitive 
requests. In the centralized one, \emph{Prophecy}, the \emph{Sketcher} 
completely avoids BFT access but now becomes a centralized component.

Amir et al.~\cite{Amir05steward:scaling} introduce \emph{Steward}, a hierarchical 
BFT architecture, that tries to scale BFT to a wide-area network, 
by introducing an abstraction layer above PBFT using a Paxos-based protocol. 
It uses a threshold signature scheme to ensure the recipient of a cross-domain 
message that enough replicas at the originating site agreed with the request. 
Both of these features are welcome to security-conscious Internet application services.
Unfortunately, no source code is readily available.

Vandiver et al.~\cite{vandiver-sosp-2007} and Garcia et al.~\cite{garcia-eurosys-2011} introduce 
middleware for BFT database replication.  Incoroporating legacy databases into a BFT system
is important for a wide range of Internet applications.  Unfortunately, both systems assume
closed systems with a finite number of clients.  The developer of an Internet-facing application
service still must deal with the issue of having end-user clients issue requests to the replicated
database system.  Either these systems need to provide support for dynamic client management or they
must offload the Internet-facing application component accepting customer/user requests to a centralized
component, something not appropriate for our particular application.

Finally, Guerraoui et al.~\cite{guerraoui-eurosys-2010} introduce a new abstraction allowing for 
the construction of new BFT protocols with a fraction of the code currently necessary, thus vastly
simplifying the BFT researcher's task.  Having waded through the 20,000 lines of PBFT code, we
applaud this effort and emphasize here the need to simplify the end application developer's task as
well.

\section{Conclusion}
\label{sect:concl}


This paper is a call to the systems community to look more closely at BFT from the perspective
of a real-world application developer.  Our experience in trying to apply the PBFT approach
to a real-world application with stringent security and reliability needs reveals 
a slew of difficulties that the application developer must face if he wants to use 
even the mature, stable and well-tuned PBFT protocol and codebase upon which a large
majority of subsequent BFT systems is based.
While the difficulties encountered by the developer can be overcome, they require
significant engineering effort and have unclear performance ramifications.  These
two characteristics are likely to make the developer hesitant to invest the effort to
leverage BFT techniques.

The systems community prides itself on building and measuring real systems.  If BFT systems
are to see widepread deployment in real-world systems, then 
the research community needs to focus on the \emph{usability} of BFT algorithms
for real world applications, from the end-developer perspective, in addition
to continuing to improve BFT middleware performance, robustness, and
deployment layouts.


Interestingly, we may find that the current BFT debate may evolve to resemble
the microkernel debate~\cite{lietde-1995}, with one camp advocating that the BFT
concept is ultimately impractical for real-world applications~\cite{birman-rds-book} and the other camp
advocating that it is not the concept that is impractical/faulty, but it is the implementation that is
impractical/faulty.  Building a \emph{complete} implementation that supports a real application 
for a long duration rather than for the length of time it takes to build
and test a prototype implementation, that does not cut corners, that is not missing features,
that does not make optimizations that break down in corner cases, that can be applied to more than
one application, and that has good performance will go a long way to settling the debate.
A tall order, for sure.

\bibliography{MyBibFile}

\begin{thebibliography}{33}
\providecommand{\natexlab}[1]{#1}
\providecommand{\url}[1]{\texttt{#1}}
\expandafter\ifx\csname urlstyle\endcsname\relax
  \providecommand{\doi}[1]{doi: #1}\else
  \providecommand{\doi}{doi: \begingroup \urlstyle{rm}\Url}\fi

\bibitem[SQL()]{SQLite/site}
Sqlite embedded database engine.
\newblock \url{http://www.sqlite.org }.

\bibitem[upR()]{upRight-project}
upright: Making distributed systems up (available) and right (correct).
\newblock \url{http://code.google.com/p/upright/w/list}.

\bibitem[mer(1987)]{merkle87}
A digital signature based on a conventional encryption function.
\newblock In \emph{CRYPTO}, 1987.

\bibitem[Abd-El-Malek et~al.(2005)Abd-El-Malek, Ganger, Goodson, Reiter, and
  Wylie]{abd-el-malek-sosp-2005}
M.~Abd-El-Malek, G.~Ganger, G.~Goodson, M.~Reiter, and J.~Wylie.
\newblock Fault-scalable byzantine fault-tolerant services.
\newblock In \emph{SOSP}, October 2005.

\bibitem[Amir et~al.(2006)Amir, Danilov, Dolev, Kirsch, Lane, Nita-rotaru,
  Olsen, and Zage]{Amir05steward:scaling}
Y.~Amir, C.~Danilov, D.~Dolev, J.~Kirsch, J.~Lane, C.~Nita-rotaru, J.~Olsen,
  and D.~Zage.
\newblock Steward: Scaling byzantine fault-tolerant systems to wide area
  networks.
\newblock In \emph{DSN}, 2006.

\bibitem[Amir et~al.(2008)Amir, Coan, Kirsch, and Lane]{amir-dsn-2008}
Y.~Amir, B.~Coan, J.~Kirsch, and J.~Lane.
\newblock Byzantine replication under attack.
\newblock In \emph{DSN}, Jun 2008.

\bibitem[Birman(2005)]{birman-rds-book}
K.~P. Birman.
\newblock \emph{Reliable Distributed Systems}.
\newblock Springer, first edition, 2005.

\bibitem[Castro and Liskov(1999)]{castro-osdi-1999}
M.~Castro and B.~Liskov.
\newblock Practical byzantine fault tolerance.
\newblock In \emph{OSDI}, February 1999.

\bibitem[Castro et~al.(2003)Castro, Rodrigues, and Liskov]{castro03base}
M.~Castro, R.~Rodrigues, and B.~Liskov.
\newblock {BASE}: Using abstraction to improve fault tolerance.
\newblock \emph{ACM TOCS}, 21\penalty0 (3), Aug. 2003.

\bibitem[Clement et~al.(2009{\natexlab{a}})Clement, Kapritsos, Lee, Wang,
  Alvisi, Dahlin, and Riche]{clement-sosp-2009}
A.~Clement, M.~Kapritsos, S.~Lee, Y.~Wang, L.~Alvisi, M.~Dahlin, and T.~Riche.
\newblock Upright cluster services.
\newblock In \emph{SOSP}, Oct 2009{\natexlab{a}}.

\bibitem[Clement et~al.(2009{\natexlab{b}})Clement, Wong, Alvisi, and
  Dahlin]{clement-nsdi-2009}
A.~Clement, E.~Wong, L.~Alvisi, and M.~Dahlin.
\newblock Making byzantine fault tolerant systems tolerate byzantine faults.
\newblock In \emph{NSDI}, April 2009{\natexlab{b}}.

\bibitem[Cowling et~al.(2006)Cowling, Myers, Liskov, Rodrigues, and
  Shrira]{cowling-osdi-2006}
J.~Cowling, D.~Myers, B.~Liskov, R.~Rodrigues, and L.~Shrira.
\newblock Hq relication: A hybrid quorum protocol for byzantine fault
  tolerance.
\newblock In \emph{OSDI}, Nov 2006.

\bibitem[Desmedt and Frankel(1989)]{DesFra89}
Desmedt and Frankel.
\newblock Threshold cryptosystems.
\newblock In \emph{CRYPTO: Proceedings of Crypto}, 1989.

\bibitem[Distler and Kapitza(2011)]{distler-eurosys-2011}
T.~Distler and R.~Kapitza.
\newblock Increasing performance in byzantine fault-tolerant systems with
  on-demand replica consistency.
\newblock In \emph{EuroSys}, Apr 2011.

\bibitem[Distler et~al.(2011)Distler, Kapitza, Popov, Reiser, and
  Schroder-Preikschat]{distler-ndss-2011}
T.~Distler, R.~Kapitza, I.~Popov, H.~Reiser, and W.~Schroder-Preikschat.
\newblock Spare: Replicas on hold.
\newblock In \emph{NDSS}, Feb 2011.

\bibitem[Garcia et~al.(2011)Garcia, Rodrigues, and
  Preguica]{garcia-eurosys-2011}
R.~Garcia, R.~Rodrigues, and N.~Preguica.
\newblock Efficient middleware for byzantine fault tolerant database
  replication.
\newblock In \emph{EuroSys}, Apr 2011.

\bibitem[Guerraoui et~al.(2010)Guerraoui, Knezevic, Quema, and
  Vukolic]{guerraoui-eurosys-2010}
R.~Guerraoui, N.~Knezevic, V.~Quema, and M.~Vukolic.
\newblock The next 700 bft protocols.
\newblock In \emph{EuroSys}, Apr 2010.

\bibitem[Herlihy and Wing(1990)]{TOPLAS:HerlihyW1990}
M.~Herlihy and J.~M. Wing.
\newblock Linearizability: {A} correctness condition for concurrent objects.
\newblock \emph{ACM TPLS}, 12\penalty0 (3):\penalty0 463--492, July 1990.

\bibitem[Kiayias et~al.(2006)Kiayias, Korman, and Walluck]{kiayias-acsac-2006}
A.~Kiayias, M.~Korman, and D.~Walluck.
\newblock An internet voting system supporting user privacy.
\newblock In \emph{ACSAC}, Dec 2006.

\bibitem[Kotla and Dahlin(2004)]{kotla-dsn-2004}
R.~Kotla and M.~Dahlin.
\newblock High throughput byzantine fault tolerance.
\newblock In \emph{DSN}, Jun 2004.

\bibitem[Kotla et~al.(2007)Kotla, Alvisi, Dahlin, Clement, and
  Wong]{kotla-sosp-2007}
R.~Kotla, L.~Alvisi, M.~Dahlin, A.~Clement, and E.~Wong.
\newblock Zyzzyva: Speculative byzantine fault tolerance.
\newblock In \emph{SOSP}, Oct 2007.

\bibitem[Lamport(1978)]{Lamport78}
L.~Lamport.
\newblock The implementation of reliable distributed multiprocess systems.
\newblock \emph{Computer Networks}, 2, 1978.

\bibitem[Lamport et~al.(1982)Lamport, Shostak, and Pease]{lamport-tpls-1982}
L.~Lamport, R.~Shostak, and M.~Pease.
\newblock The byzantine generals problem.
\newblock \emph{ACM TPLS}, 4\penalty0 (3):\penalty0 382--401, July 1982.

\bibitem[Liedtke(1995)]{lietde-1995}
J.~Liedtke.
\newblock On micro-kernel construction.
\newblock \emph{ACM SIGOPS Operating Systems Review}, 29\penalty0 (5), Dec
  1995.

\bibitem[Lynch(1996)]{Lynch:1996:DA}
N.~Lynch.
\newblock \emph{Distributed Algorithms}.
\newblock Morgan Kaufmann, 1996.

\bibitem[Merideth et~al.(2005)Merideth, Iyengar, Mikalsen, Tai, Rouvellou, and
  Narasimhan]{Merideth05}
M.~Merideth, A.~Iyengar, T.~Mikalsen, S.~Tai, I.~Rouvellou, and P.~Narasimhan.
\newblock Thema: Byzantine-fault-tolerant middleware for web-service
  applications.
\newblock In \emph{SRDS}, Oct. 2005.

\bibitem[Pallemulle et~al.(2008)Pallemulle, Thorvaldsson, and
  Goldman]{Pallemulle08}
S.~L. Pallemulle, H.~D. Thorvaldsson, and K.~J. Goldman.
\newblock Byzantine fault-tolerant web services for n-tier and service oriented
  architectures.
\newblock In \emph{ICDCS}, June 2008.

\bibitem[Schneider(1990)]{schneider-computingSurv-1990}
F.~Schneider.
\newblock Implementing fault-tolerant services using the state machine
  approach: a tutorial.
\newblock \emph{ACM Computing Surveys}, 22\penalty0 (4):\penalty0 299--319, Dec
  1990.

\bibitem[Sen et~al.(2010)Sen, Lloyed, and Freedman]{sen-nsdi-2010}
S.~Sen, W.~Lloyed, and M.~Freedman.
\newblock Prophecy: Using history for high-throughput fault tolerance.
\newblock In \emph{NSDI}, April 2010.

\bibitem[Singh et~al.(2008)Singh, Das, Maniatis, Druschel, and
  Roscoe]{conf/nsdi/SinghDMDR08}
A.~Singh, T.~Das, P.~Maniatis, P.~Druschel, and T.~Roscoe.
\newblock {BFT} protocols under fire.
\newblock In \emph{NSDI}, 2008.

\bibitem[Vandiver et~al.(2007)Vandiver, Balakrishnan, Liskov, and
  Madden]{vandiver-sosp-2007}
B.~Vandiver, H.~Balakrishnan, B.~Liskov, and S.~Madden.
\newblock Tolerating byzantine faults in transaction processing systems using
  commit barrier scheduling.
\newblock In \emph{SOSP}, Oct 2007.

\bibitem[Wood et~al.(2011)Wood, Singh, Venkataramani, Shenoy, and
  Cecchet]{wood-eurosys-2011}
T.~Wood, R.~Singh, A.~Venkataramani, P.~Shenoy, and E.~Cecchet.
\newblock Zz and the art of practical bft.
\newblock In \emph{EuroSys}, April 2011.

\bibitem[Yin et~al.(2003)Yin, Martin, Venkataramani, and adn
  M.~Dahlin]{yin-sosp-2003}
J.~Yin, J.-P. Martin, A.~Venkataramani, and L.~A. adn M.~Dahlin.
\newblock Separating agreement from execution for byzantine fault tolerant
  services.
\newblock In \emph{SOSP}, Oct 2003.

\end{thebibliography}
\bibliographystyle{abbrvnat}





\end{document}